\documentclass[aps,prl,twocolumn,nofootinbib,longbibliography]{revtex4-1}
\usepackage{amsmath}
\usepackage{amssymb}
\usepackage{graphicx}
\usepackage{color}
\usepackage{mathrsfs}

\graphicspath{{./Figures/}}

\renewcommand*{\emptyset}{\varnothing}
\DeclareMathOperator{\Tr}{Tr}

\newcommand*{\scri}{\mathscr{I}} % Null infinity
\newcommand*{\swap}{\mathcal{S}} %Swap operator
\newcommand*{\island}{\mathcal{I}}

\begin{document}

\title{\Large The Page curve and baby universes}
%\date{\today}
\author{Donald Marolf$^a$}
\author{Henry Maxfield$^b$}
\affiliation{Department of Physics, University of California, Santa Barbara, CA 93106, USA \\
$^a$marolf@ucsb.edu, $^b$hmaxfield@physics.ucsb.edu}
\begin{abstract}
Black hole thermodynamics suggests that,
in order to describe the physics of distant observers, one may model a black hole as a standard  quantum system with density of states set by the Bekenstein-Hawking entropy $S_\mathrm{BH}$.  This idea has long been considered to be in strong tension with Hawking's prediction that radiation from black holes is nearly thermal, and with low-energy gravity more generally.  But the past two years have shown that low-energy gravity \emph{does} offer a self-consistent description of black hole evaporation consistent with the above idea, and which in particular reproduces the famous Page curve.  We provide a brief overview of this new paradigm% aimed at researchers who have not yet digested it in full
, focusing on Lorentz-signature asymptotically-flat spacetimes, and emphasizing operationally-defined observables that probe the entropy of Hawking radiation.
\end{abstract}

\maketitle

\section{Introduction}

The past two years have witnessed a startling transformation in discussions of black hole information.
Before 2019, low-energy gravitational physics appeared incontrovertibly to lead to information loss in the evaporation of black holes, raising questions about pair-production rates for `remnants' of black hole evaporation, the status of the first and second laws of black hole thermodynamics, and consistency with string theory and AdS/CFT.  These tensions suggested to many physicists that black holes might instead enjoy a feature which we will call `Bekenstein-Hawking unitarity' (BH unitarity): in order to describe the physics of distant observers, one may model a black hole as a standard  quantum system with density of states set by the Bekenstein-Hawking entropy $S_\mathrm{BH}$, with unitary evolution (up to couplings with its environment). But BH unitarity appeared to require truly new physics coming from some complete formulation of quantum gravity and which was not at all apparent in the low-energy theory. See e.g.\ \cite{Mathur:2009hf,Harlow:2014yka,Unruh:2017uaw,Marolf:2017jkr}
for reviews.

\begin{figure}
	\centering
	\includegraphics[width=.7\columnwidth]{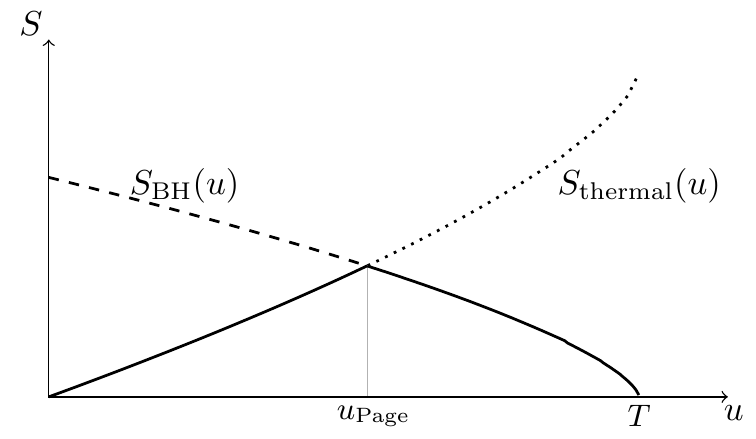}
	\caption{The Page curve for the entropy of Hawking radiation emitted before time $u$ (solid curve).
In Hawking's original calculation the entropy of radiation emitted during evaporation is always equal to the radiation's thermal entropy $S_\mathrm{thermal}$ (the always-increasing dotted curve).  But BH unitarity predicts that operational measurements of radiation emitted during evaporation will be consistent with a density of states set by $S_\mathrm{BH}\sim \frac{A}{4G}$ (dashed curve), and since to good approximation the emitted radiation is entangled only with the black hole this $S_\mathrm{BH}$ must also bound measurements of the radiation's entropy. So for BH unitarity to hold, as described by Page such measurements should lie close to the solid curve defined at each retarded time $u$ by the lesser of $S_\mathrm{thermal}, S_\mathrm{BH}$.
%For a while this is increasing, given by the thermal entropy (dotted curve). But under BH unitarity it is bounded by the Bekenstein Hawking entropy $S_{BH}\sim \frac{A}{4G}$  Consequently, after the Page time $u_\mathrm{Page}$ the entropy must decrease, approximately saturating that bound.
\label{fig:Page}}
\end{figure}

In striking contrast,
it is now clear that low-energy gravity \emph{does} in fact offer a self-consistent description of black hole evaporation in which BH unitarity holds \cite{Penington:2019npb,Almheiri:2019psf,Penington:2019kki,Almheiri:2019qdq,Marolf:2020xie,Almheiri:2020cfm,Marolf:2020rpm}.
The key piece of evidence is a semiclassical calculation of operationally-defined measurements of the entropy of radiation, involving `quantum-extremal islands' and `replica wormholes'. This calculation reproduces the `Page curve' shown in figure \ref{fig:Page}, long regarded as a smoking gun for BH unitarity. As a result, the outcomes of any operational tests performed on the radiation (as predicted by low-energy gravity) are consistent with unitary evolution and density of states $e^{S_\mathrm{BH}}$.
Nonetheless, in this description Hawking's original information loss calculation still appears to be fundamentally correct, so that the von Neumann entropy of the resulting density matrix does \emph{not} follow the Page curve. The consistency of these two facts relies on ideas from the late 1980's involving spacetime wormholes and baby universes.

This note provides a brief overview of the new paradigm, summarising the central ideas; we refer to \cite{Marolf:2020rpm} for more complete details and references to the literature; see also the recent review \cite{Almheiri:2020cfm}. We will present these ideas in an asymptotically flat setting (though the conclusions are no different with AdS asymptotics), formulating gravity in terms of a path integral over Lorentz-signature metrics, and emphasizing the operational tests to which we referred above.

%From this perspective, while low-energy gravity makes predictions in line with BH unitarity, it does not give definite predictions for the detailed dynamics.
% Instead, there are a plethora of superselection sectors for distant observers associated with the state of baby universes, each sector giving

%In striking contrast,
%it is now clear that low-energy gravity \emph{does} in fact offer a self-consistent description of black hole evaporation in which BH unitarity holds. Nonetheless, in this description Hawking's original information loss calculation still appears to be fundamentally correct, leading to a formal density matrix for the emitted radiation which is highly mixed. Yet any operational tests performed on the radiation are consistent with unitary evolution and density of states $e^{S_\mathrm{BH}}$.  The new picture combines old ideas from the late 1980's involving baby universes and spacetime wormholes with a new understanding of semiclassical contributions associated with quantum-extremal islands and replica wormholes.  In particular, as we stress below, the latter provide a semiclassical calculation showing that operational measurements of the entropy of radiation emitted during black hole evaporation follow the `Page curve' predicted by BH unitarity and shown in figure  \ref{fig:Page}.

%\textbf{Bekenstein-Hawking unitarity:} in order to describe measurements of distant observers, black holes can be modelled as a quantum system with density of states $e^{S_\mathrm{BH}}$ whose evolution is unitary.

\section{Summary}

Entropy is not a directly observable quantity for a single system: it can only be inferred from measurements performed on many copies of a system. We will begin by introducing the `swap entropy', an entropy-like quantity defined in terms of such measurements. This is constructed to coincide with the von Neumann entropy when we are given many uncorrelated copies of a state. However, our later calculations will involve spacetime wormholes, which are  %generally defined as
 shortcuts between otherwise well-separated regions of spacetime and which for us will connect the interiors of distant black holes. Such connections induce unavoidable correlations between distant experiments, even those performed in otherwise separate universes. As a result, taking such wormholes into account one finds that the states of Hawking radiation emitted from identical black holes will not in fact be independent.  This then allows the swap entropy to differ from the von Neumann entropy.

%  Our presentation broadly follows  in terms of notation and terminology, and in particular using the same asymptotically-flat Lorentz-signature setting (see also XXXX --\DM{how to treat other asymptotically flat papers or those that at least mention Lorentz-signature?}).

%     As was discussed already in the 1980's, a key physical point is that semiclassical quantum gravity can include contributions from so-called `spacetime wormholes', which are shortcuts between what would otherwise be well-separated region of spacetime.  They thus lead directly to failures of cherished principles like the idea (closely related to cluster-decomposition) that well-separated experiments may be treated as being independent.  Indeed, such wormholes mean that even experiments performed in otherwise-separate universes can be strongly correlated.  We thus begin by reviewing an entropy-like quantity that we call `swap entropy'  $S_{swap}$ which coincides with the formal von Neumann entropy when the above long-range correlations can be avoided, and whose difference from the formal von Neumann entropy can thus be used to detect such novel correlations when they exist.

After summarising the framework of semiclassical gravity to be used, we then briefly describe recent computations of such swap entropies for radiation emitted from evaporating black holes. While the formal von Neumann entropy associated with a single black hole receives contributions only from a single saddle-point (the `Hawking saddle'), the swap entropies also receive contributions from replica wormholes.  As a result, with standard initial conditions the formal von Neumann entropy  $S_\mathrm{vN}$ of radiation emitted by a single black hole is indeed thermal ($S_\mathrm{vN} = S_\mathrm{thermal}$), while the swap entropy $S^\mathrm{swap}$ follows the Page curve (solid curve in figure \ref{fig:Page}).

Finally, we explain the physics associated with the discrepancy between $S_\mathrm{vN}$ and $S^\mathrm{swap}$ in terms of entanglement with baby universes.  This entanglement mediates strong correlations between the end products of distinct experiments that form and evaporate otherwise-identical black holes, even when the experiments are well-separated in space and time.  However, for robust reasons, the correlations between any set of evaporation experiments must be entirely classical.  In particular, one may describe these correlations by declaring that identical evaporation experiments always lead to identical pure states of Hawking radiation, but that the pure state is drawn at random from an appropriate ensemble.  But crucially, for a given process that forms and evaporates black holes, all iterations of the process are governed by the results of a \emph{single} random draw no matter how many times the experiment is performed.  This basic picture was described long ago \cite{Coleman:1988cy,Giddings:1988cx,Giddings:1988wv,Polchinski:1994zs}. But with the replica wormhole Page curve results for $S^\mathrm{swap}$, the paradigm is now fully consistent with low-energy physics, and with a black hole density of states given by the Bekenstein-Hawking entropy.  In this way, black hole evaporation is consistent with BH unitarity, but the detailed dynamics (as predicted by low-energy gravity) is random, rather than uniquely determined.

\section{Swap entropies}

We have already stressed that we are most interested in operational measurements performed by asymptotic observers. In general, an observer may form and evaporate some number $n$ of black holes in widely separated regions of spacetime, and make joint measurements on the resulting $n$ sets of Hawking radiation which we will call `replicas'. The predictions for such experiments are encoded in the expectation values $\Tr(\mathcal{O}\rho^{(n)})$ of operators $\mathcal{O}$ acting on the density matrix $\rho^{(n)}$ of the full $n$-replica system. In particular, $\mathcal{O}$ need not factorise into a product of operators acting on each individual replica.

If the $n$ processes of black hole formation and evaporation are identical and uncorrelated, then $\rho^{(n)}$ will be the $n$-fold tensor product of a single-replica state $\rho$: $\rho^{(n)}=\rho^{\otimes n}$.  Under such circumstances we can use observations on the $n$-replica state $\rho^{(n)}=\rho^{\otimes n}$ to determine the degree to which the single-copy sate $\rho$ is mixed.

%We formulated the information paradox as a statement about the entropy of Hawking radiation. However, our philosophy here is to concentrate on predictions for observables, and the von Neumann entropy of a state is not a directly observable quantity, since it does not depend linearly on the density matrix. How, then, would an observer distinguish between the two scenarios in figure \HM{ref}, one where the entropy increases for all time, and one where it follows the Page curve?

%The answer is that she must perform experiments on multiple copies, or `replicas', of the system. The predictions for such experiments are encoded in the expectation values $\Tr(\mathcal{O}\rho^{(n)})$ of operators $\mathcal{O}$ acting on the density matrix $\rho^{(n)}$ of the $n$-replica system. If we prepare $n$ uncorrelated states, this will be the $n$-fold tensor product of the original state, $\rho^{(n)}=\rho^{\otimes n}$.

A simple and useful such observation is the `swap test' \cite{buhrman2001quantum,Hayden:2007cs}  for $n=2$, a measurement of the swap operator $\swap$ that acts by exchanging the two replicas: $\swap |\psi_1\rangle \otimes |\psi_2\rangle = |\psi_2\rangle \otimes |\psi_1\rangle$. More generally, on $n$ replicas we can measure the operator $U_\tau$ which enacts the cyclic permutation $\tau = (1\ 2\cdots n)$ on the replicas.\footnote{For $n>2$, $U_\tau$ is not Hermitian, but it is normal (commutes with its adjoint) so can be measured.} We may encode the expectation values $\Tr\left( U_\tau \rho^{(n)} \right)$ of these permutation operators in `swap entropies' $S_n^{\text{swap}}$, defined by
\begin{equation}
\label{eq:SwapEntn}
S_n^{\text{swap}}(\rho^{(n)}) : = -\frac{1}{n-1} \log \Tr \left( U_\tau \rho^{(n)} \right).
\end{equation}
The reason for this name arises from the uncorrelated-but-identical case $\rho^{(n)}=\rho^{\otimes n}$ where $\rho^{(n)}$ consists of $n$ uncorrelated copies of a state $\rho$. Then, $S_n^{\text{swap}}$ reduces to the R\'enyi entropy defined by
\begin{equation}
\label{eq:Renyi2}
S_n(\rho) : = -\frac{1}{n-1} \log \Tr \left(\rho^n \right).
%S(\rho) := -\Tr\left(\rho\log\rho\right) = \lim_{n\to 1} S_n(\rho) \label{eq:vN}
\end{equation}
We may thus summarize this result by writing
\begin{equation}
\label{eq:uncorrelatedSwap}
\rho^{(n)}=\rho^{\otimes n} \implies S_n^{\text{swap}}(\rho) =S_n(\rho).
\end{equation}

However, when we study the state of radiation $\rho^{(n)}$ from $n$ separate black holes below, we will find reason to believe that $\rho^{(n)}$ is very different from $\rho^{\otimes n}$, and in particular that it contains strong correlations between the various sets of radiation collected. Then $S_n^{\text{swap}}$ need not equal $S_n$, but will contain new physics that we will interpret more fully below.

Finally, we will also find it convenient to consider the $n\to 1$ limit of the swap entropy $S^{\text{swap}} : = \lim_{n\to 1} S_n^{\text{swap}}$, in analogy with the representation of von Neumann entropy as the $n\rightarrow 1$ limit of R\'enyis:
\begin{equation}
S(\rho) := -\Tr\left(\rho\log\rho\right) = \lim_{n\to 1} S_n(\rho) \label{eq:vN}.
\end{equation}
For swap entropies this limit is somewhat formal\footnote{Though in many contexts it can be defined by analytic continuation using Carlson's theorem from complex analysis, by using the resolvent method of \cite{Penington:2019kki}, or by using the methods of \cite{DHoker:2020bcv}.}, since the swap entropy is strictly speaking only defined for integers $n\geq 2$. Nonetheless, there will be a natural way to formulate the calculation of the swap entropy so that it applies at general real $n$, and it will simplify in the $n\to 1$ limit. We interpret the result as the von Neumann entropy that would be deduced from the observations of a sophisticated experimentalist holding the state $\rho^{(n)}$ with many replicas, after taking all experimental precautions to ensure that the replicas are as uncorrelated as is possible.\footnote{This in fact requires exponentially many copies of the system, $n\gg e^{S^{\text{swap}}}$ where $S^{\text{swap}}$ is often comparable to the density of states of some controlling subsystem.}

The above construction may be interpreted as a physical version of the so-called `replica trick' often used to compute the von Neumann entropy of a density matrix $\rho$.  But our version is not a trick in any way.  Each replica is a physical system on which we can perform joint measurements.  And potential correlations between our replicas are not specified or forbidden by hand.    Instead,  such correlations are determined by the dynamics and initial conditions.

\section{A first semiclassical contribution:  the Hawking saddle}

Now that we have described the observables of interest, we proceed to lay out the framework we will use to calculate their expectation values.

We will consider states $\rho^{(n)}(u)$ defined by preparing $n$ identical initial pure states of matter destined to collapse into black holes, allowing them to evaporate, and collecting the Hawking radiation up to some time $u$. If these $n$ black holes are sufficiently widely separated in space or time, locality (cluster decomposition) would suggest that the $n$ states are uncorrelated: we will see later that this expectation needs to be modified, which is why we are belabouring the distinction between von Neumann entropy and measurable quantities. We will idealise this distant separation by using $n$ separate asymptotic boundaries, preparing the initial collapsing matter state on $n$ separate copies of $\scri^-$, and collecting Hawking radiation on $n$ corresponding copies of $\scri^+$ (up to retarded time $u$).

Let us first consider the case $n=1$ of a single formation-evaporation experiment. With a single black hole, we may compute the expectation values of any operator $\mathcal{O}$ (supported on $\scri^+$ before time $u$) measuring the Hawking radiation. The knowledge of all such expectation values is equivalent to the $n=1$ density matrix $\rho^{(1)}(u)$ of the Hawking radiation, with expectation values given by $\Tr(\rho^{(1)}(u)\mathcal{O})$.  To compute these expectation values, we use the in-in formalism in which the pure initial state $|\mathrm{in} \rangle$ is described by the density matrix $\rho_\mathrm{in} = |\mathrm{in} \rangle \langle\mathrm{in} |$, which involves both a bra and a ket.  The path integral then evolves both the bra and the ket into the future in what one might call separate copies of the spacetime, for the moment in fact defining two separate path integrals.  We take the evolution to stop at some achronal surface $\Sigma_u$ such that it would form a Cauchy surface when taken together with the points on $\scri^+$ before retarded time $u$.  We sew the two path integrals together along $\Sigma_u$ in order to trace over states that remain in the spacetime and have not yet escaped to $\scri^+$ before the retarded time $u$. Finally, we sum over states on the portion of $\scri^+$ before time $u$, weighted according to the matrix elements of the operator $\mathcal{O}$ of interest.\footnote{One might like to compute the matrix elements of the density matrix $\rho^{(1)}(u)$ directly, by choosing $\mathcal{O} = |j\rangle\langle i|$, where $|i\rangle$ runs over some complete basis of states on the relevant portion of $\scri^+$. We avoid this, because all such states would have enormous energy at retarded time $u$ due to the absence of entanglement present in any state close to the vacuum, rendering the calculation beyond semiclassical control.}

%One might like to compute matrix elements ${}_u\langle i| \rho^{(1)}(u)|j\rangle_u$ of the $n=1$ density matrix between pure states $|i \rangle_u$ and  $|j \rangle_u$.  Such matrix elements are associated with the path integral depicted in figure \ref{fig:rho1Hawk}.  Here we use the so-called in-in formalism in which the pure initial state $|in \rangle$ is described by the density matrix $\rho_{in} = |in \rangle \langle in |$, which involves both a bra and a ket.  The path integral then evolves both the bra and the ket into the future in what one might call separate copies of the spacetime, for the moment in fact defining two separate path integrals.  We take the evolution to stop at some achronal surface $\Sigma_u$ such that it would form a Cauchy surface when taken together with the points on $\scri^+$ before retarded time $u$.  Finally, we sew the two path integrals together along $\Sigma_u$ in order to trace over states that remain in the spacetime and have not yet escaped to $\scri^+$ before the retarded time $u$.

\begin{figure}
	\centering
	\includegraphics[width=.7\columnwidth]{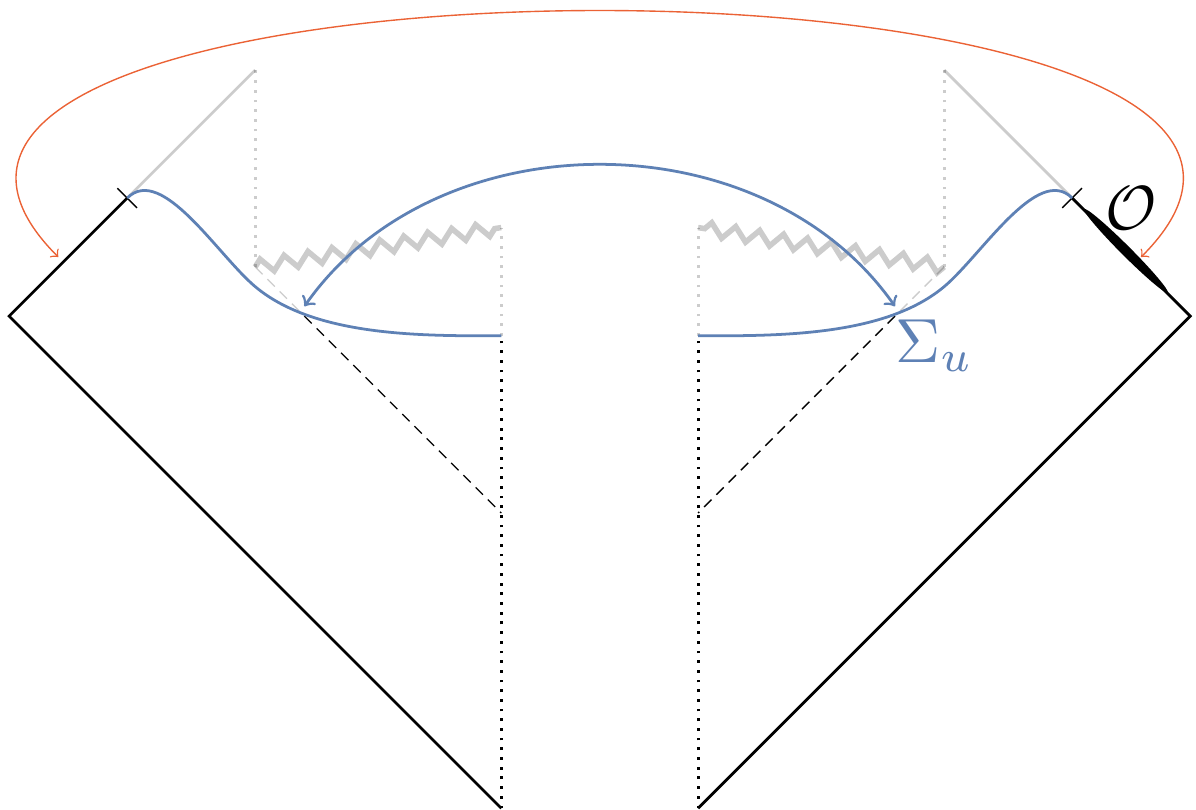}
	\caption{An artists impression of the path integral computation of $\Tr(\mathcal{O} \rho^{(1)}(u))$.  The bra and ket copies of the spacetime pieces are sewn together along the blue partial Cauchy surface $\Sigma_u$ which meets $\scri^+$ at retarded time $u$, and along $\scri^+$ before that time after acting with the operator $\mathcal{O}$ there.}
\label{fig:rho1Hawk}
\end{figure}

It is useful that the diagram in figure \ref{fig:rho1Hawk} can be read at several levels.  First, one could take it to describe a calculation in quantum field theory (including perturbative gravitons) on a spacetime with a fixed metric that describes a collapsing black hole as in Hawking's original calculation \cite{Hawking:1974sw}.  Second, one could take it to describe a similar calculation in which the spacetime metric is chosen to describe a black hole that first collapses and then evaporates, with a black hole that slowly decays by radiating energy to infinity according to semi-classical expectations.  In such a spacetime the black hole mass eventually shrinks down to become Planck scale at some finite retarded time $u_\mathrm{Planck}$, after which the spacetime will contain Planck-scale curvatures whose physics is not under semi-classical control.  But such concerns are avoided by our considering only the portion of $\scri^+$ before some finite retarded time $u$ so long as we choose $u$ to be significantly less than $u_\mathrm{Planck}$. In either case, we integrate over quantum fields with the stated boundary conditions at $\scri^-$ and $\scri^+$, and which satisfy the desired sewing condition on the bulk partial Cauchy surface shown in blue.

Finally, we can take figure \ref{fig:rho1Hawk} to describe semiclassical contributions to the full quantum gravity description of our process.  In practice,  this means that the metric must solve the semi-classical Einstein equations. These equations are sourced by a quantum stress-tensor which includes the effect of integrating out fields at one loop, to incorporate the back-reaction from Hawking radiation. To be more precise, one might like to integrate over a family of such geometries to incorporate quantum fluctuations in the rate of radiation of energy, momentum and angular momentum.% as emphasised recently by \cite{Flanagan:2021ojq}.

%At this stage we should admit that our choice to compute particular matrix elements of $\rho^{(1)}(u)$ presents a technical problem, in that taking the quantum state to be pure on the {\it portion} of $\scri^+$ before retarded time $u$ means that there are no correlations between $\scri^+$ and the blue partial Cauchy surface.  Any quantum state of that sort would of course differ dramatically at retarded time $u$ from the field theory vacuum, and would thus be expected to lead to enormous stress-energies at $u$ whose back-reaction will fail to be under semiclassical control.  However, this technical issue is easily deal with by summing over classes of boundary conditions at $\scri^+$ that may differ markedly just before the chosen retarded time $u$ but which coincide at earlier retarded times; i.e., instead of computing ${}_u\langle i| \rho^{(1)}(u)|j\rangle_u$, we compute
%$\Tr \left({\cal O}_{ij} \rho^{(1)}(u) \right)$, where ${\cal O}_{ij}$ is an operator on the part of $\scri^+$ before retarded time $u$ that agrees with $|j\rangle_u {}_u \langle i|$ when restricted to significantly earlier times but which allows for vacuum-like correlations between the UV degrees of freedom near $u$ and the blue partial Cauchy surface shown.  However, we will leave this technical fix implicit in our discussions below and -- for pedagogical simplicity -- we will continue to discuss matrix elements ${}_u\langle i| \rho^{(1)}(u)|j\rangle_u$ as if they did in fact receive well-defined semiclassical contributions of the form shown in figure \ref{fig:rho1Hawk}.

With any of the above interpretations, the resulting expectation values  $\Tr(\rho^{(1)}(u)\mathcal{O})$ will be much as anticipated in \cite{Hawking:1976ra}.  In particular, they will be consistent with a roughly  `thermal' state, though with a temperature that evolves adiabatically as a function of $u$ in a manner dictated by conservation of energy and the equation of state of the evaporating black hole. As a result, correlations between Hawking photons become negligible at time separations greater than the inverse of the instantaneous black hole temperature. The effects of this adiabatic variation are small and, as a result, the von Neumann entropy $S\left( \rho^{(1)}(u) \right) $ computed from such matrix elements will grow monotonically in time, with the rate at each time again being determined by the effective temperature.  For brevity, we refer to this as giving a thermal entropy $S_\mathrm{thermal}(u)$, as depicted in figure \ref{fig:Page}.

Thinking of figure \ref{fig:rho1Hawk} as describing a self-consistent solution of the semiclassical physics that satisfies the desired boundary conditions raises the question of whether there might be other such solutions. Indeed, the semiclassical description of a fully quantum process often involves finding several such solutions and summing the resulting contributions.  In a path integral description one refers to this as a sum over saddle points, and we will use this terminology below.  In particular, we will refer to contributions of the form shown in figure \ref{fig:rho1Hawk} as Hawking saddles due to their similarity to the descriptions of \cite{Hawking:1974sw} and \cite{Hawking:1976ra}.

Having described the Hawking saddle for $\rho^{(1)}(u)$, we can immediately see that taking $n$ such saddles gives a similar semiclassical contribution to expectation values for $n$ identical black holes, computed from the $n$-copy density matrix $\rho^{(n)}(u)$.  Pictorially, we simply draw $n$ separate copies (replicas) of figure \ref{fig:rho1Hawk} (perhaps with different operators $\mathcal{O}$ on the chosen portion of $\scri^+$, or more generally an operator that mixes the $n$ copies of the radiation) with no connections or identifications between replicas.

\begin{figure}
	\centering
	\includegraphics[width=.72\columnwidth]{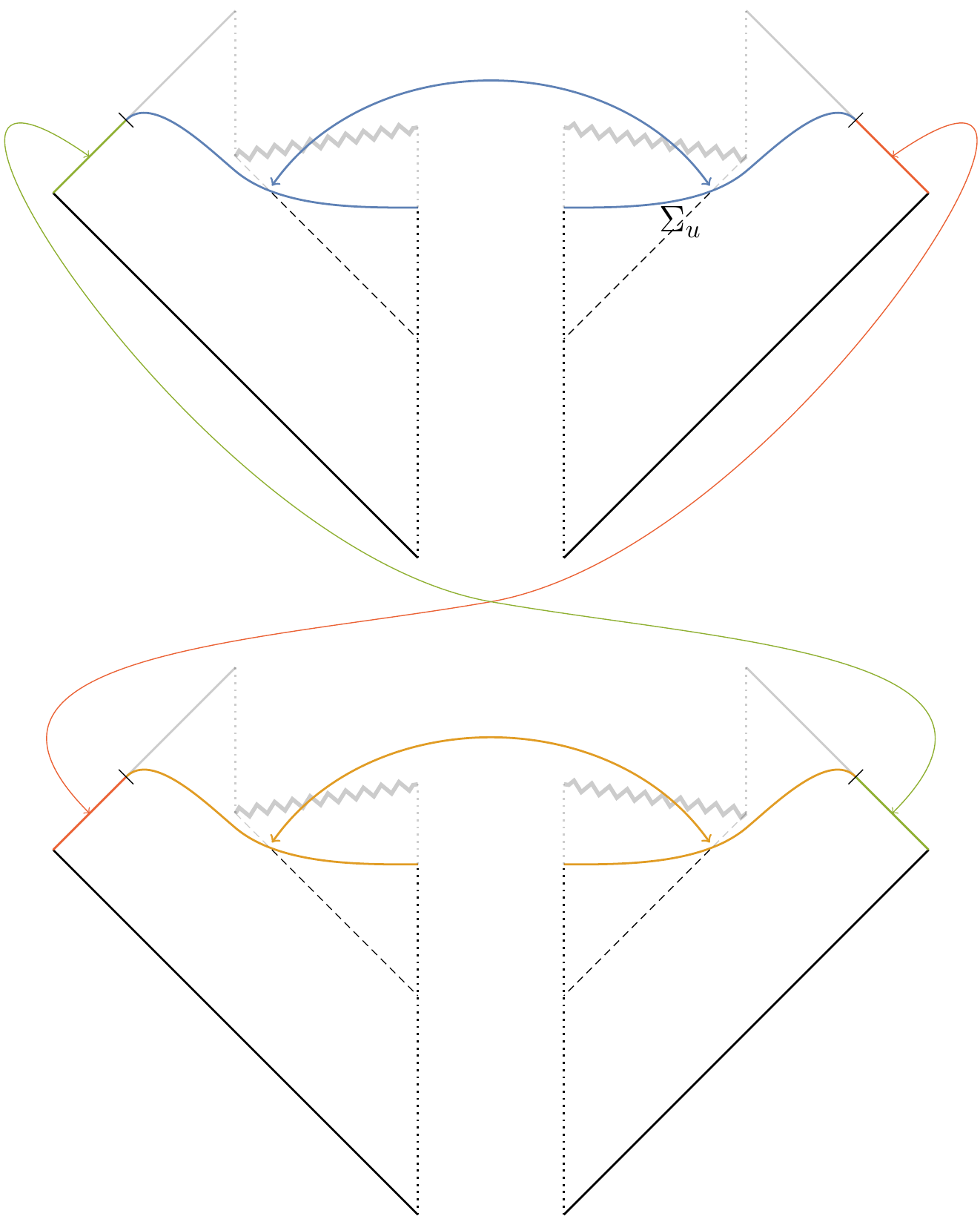}
	\caption{A saddle for $S^{\text{swap}}_2$ formed by sewing two copies of the Hawking saddle figure \ref{fig:rho1Hawk} together along the parts of $\scri^+$ before retarded time $u$.  These identifications are shown in green and red.}
\label{fig:HWforS2}
\end{figure}

Such collections of $n$ spacetimes will again be called Hawking saddles, and if these are the only contributions then $\rho^{(n)}(u)$ is just a tensor product of $n$ uncorrelated states
$\rho^{(1)}(u)$ as one would naively expect:
\begin{equation}
\label{eq:Hawkapprox}
\text{Hawking saddles give} \ \ \rho^{(n)}_{\text{Hawking}}(u) = \left(\rho^{(1)}(u) \right)^{\otimes n} .
\end{equation}
Indeed, by choosing our operator $\mathcal{O}$ to be a cyclic permutation $U_\tau$, we may consider the Hawking saddles as contributions to the swap entropies \eqref{eq:SwapEntn}.  The definition of swap entropies instructs us to sum over boundary conditions on the various copies of $\scri^+$, identifying the state on the $k$th  bra copy of $\scri^+$ with the state on $(k+1)$th ket copy as dictated by the cyclic permutation $U_\tau$ in \eqref{eq:SwapEntn}. Implementing these boundary conditions on the Hawking saddle, we arrive at a path integral as depicted in figure \ref{fig:HWforS2}.

But if we use only these Hawking saddles, the resulting swap entropy calculation is identical to a calculation of the von Neumann entropy of $\rho^{(1)}(u)$. Consistent with \eqref{eq:uncorrelatedSwap} and \eqref{eq:Hawkapprox}, if such saddles dominate we find $S^{\mathrm{swap}}_n (u) \approx S_n\left(\rho^{(1)}(u)\right)$ where the right-hand-side is just the $n$th R\'enyi entropy $S_{\mathrm{thermal}, n}(u)$ of thermal radiation at $\scri^+$ (again with appropriately adiabatically varying temperature $T(u)$).
%
%Using the above Hawking saddles to compute matrix elements of $\rho^{(n)}(u)$ then represents this swap entropy as a sum over such Hawking saddles, where we may expect that some particular configuration of quantum fields (or some class of configurations) emerges as the dominant member of this sum.  We can then depict the dominant Hawking saddle for our swap entropy as in figure \ref{fig:HWforS2} below.
  Taking the limit $n\rightarrow 1$ then yields $S^{\mathrm{swap}} \approx S_\mathrm{thermal}$, the rising dotted curve in figure \ref{fig:Page}.

\section{Replica wormholes}

We now come to the main point.  While the Hawking saddles may seem natural, the results of \cite{Penington:2019kki,Almheiri:2019qdq} strongly indicate that other saddles will also contribute to our swap entropies, and that the contributions of these new saddles will be important after the Page time.  The new saddles are known as replica wormholes.

It is useful to explain replica wormholes in the language of the gravitational path integral.  Here we take the perspective that the path integral sums over bulk spacetimes that satisfy some given set of boundary conditions dictated by the physics one wishes to study.  For example, for a fixed initial state on $\scri^-$, our $n=2$ swap entropy is associated with gluing four copies of $\scri^+$ together as shown in figure \ref{fig:HWforS2}.  The expectation value of $S^{\text{swap}}_2$ is then given by summing over bulk spacetimes compatible with the given boundary conditions.

\begin{figure}
	\centering
	\includegraphics[width=.72\columnwidth]{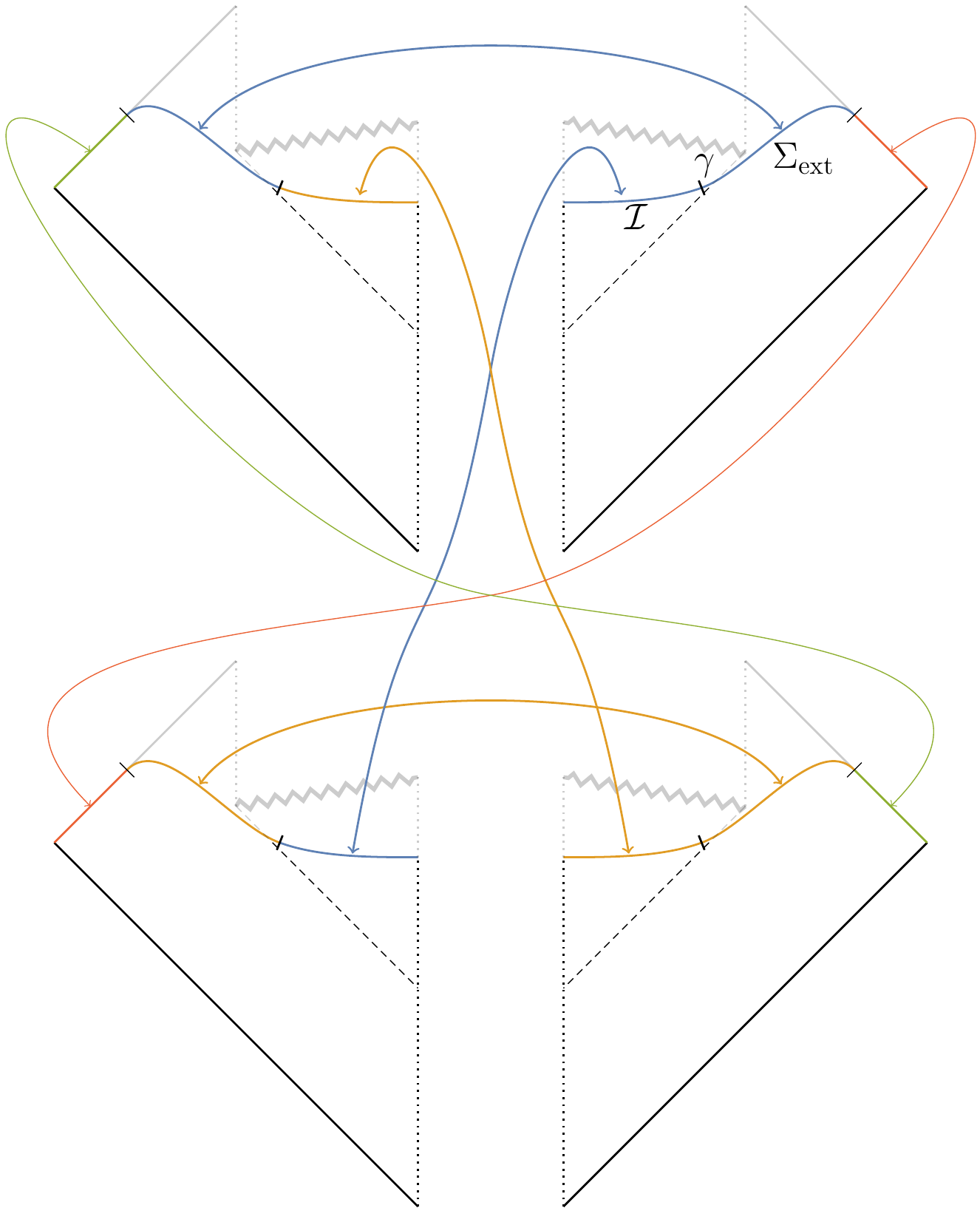}
	\caption{A replica wormhole for $S^{\text{swap}}_2$.  The boundary conditions are identical to those in figure \ref{fig:HWforS2}, but the identifications differ in the bulk due to the presence of islands (${\cal I}$). On the islands the bulk spacetimes are sewn together in the same pattern that the swap boundary conditions impose on the early part of $\scri^+$, effectively `undoing' the swap on this part of the geometry (where most of the bulk entropy resides).
}
\label{fig:RWs}
\end{figure}

Now, one can and should ask what class of bulk spacetimes should be included in this sum. For example, should we restrict to smooth spacetimes?  Or to spacetimes with appropriate causal structures? Our central assumption (motivated further below) is that this class should include `replica wormhole' spacetimes of the form depicted in figure \ref{fig:RWs}.

Such spacetimes may be constructed from the Hawking spacetimes discussed above by a `cut-and-paste' procedure, whereby we change the manner in which bra  and ket spacetimes along the future Cauchy surface $\Sigma_u$. Specifically, we may imagine starting with a geometry as in figure \ref{fig:HWforS2}, cutting each copy of $\Sigma_u$ along some so-called \emph{island} $ \island  \subset \Sigma_u$, and then sewing the replicas together cyclicly along $\island$ as shown for $n=2$ in figure \ref{fig:RWs}. That is, we choose to swap the islands in precisely the same pattern that we swap the parts of $\scri^+$ before retarded time $u$.  The result is a single connected spacetime satisfying the swap boundary conditions.

The replica wormhole geometries are locally identical to the more conventional in-in geometries considered above, except at the boundary of the island where we change the identification between replicas. This locus is a codimension-two surface $\gamma = \partial\island$, which we call the \emph{splitting surface}. Along this surface, we must introduce a new sort of singularity, which is however of a rather mild type.

\begin{figure}
	\centering
	\includegraphics[width=.6\columnwidth]{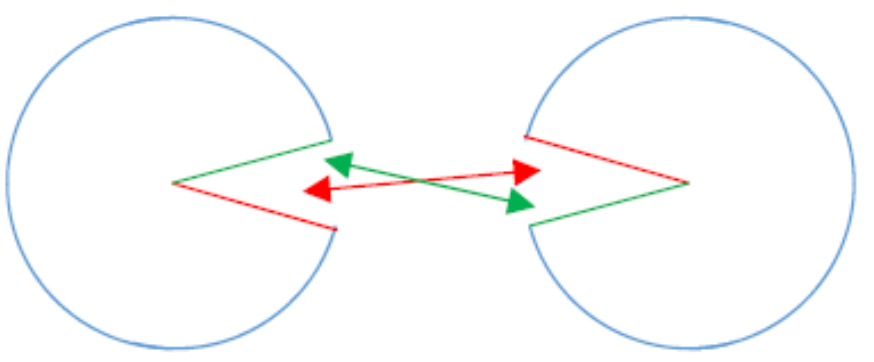}
	\caption{Making radial cuts in two disks and sewing them together with the identifications shown in red and green results in a conical excess of $2\pi$.}
\label{fig:conicalexcess}
\end{figure}

In more detail, the geometry in a neighbourhood of the splitting surface $\gamma$ is a Lorentzian analogue of a conical singularity. To explain this, it is useful to first recall certain features of conical singularities in Euclidean-signature spacetimes.  In particular, a conical excess of magnitude $2\pi (n-1)$ can be created by taking $n$ copies of a smooth spacetime, cutting them all open along a common half-line, and then pasting them together cyclicly to form a connected geometry. This process is the Euclidean version of the cutting and gluing along the island $\island$ which produced figure \ref{fig:RWs}.  The $n=2$ version of this construction in shown in figure \ref{fig:conicalexcess}.  Such an excess introduces a delta-function Ricci tensor at the conical defect which then integrates to give a finite contribution to the Euclidean Einstein-Hilbert action $\int\!\!\sqrt{g} R$.  Though we focus on Lorentz-signature spacetimes here, it is worth commenting that in Euclidean signature such conical spacetimes are just limits of smooth spacetimes in which a finite amount of Einstein-Hilbert action becomes concentrated at the defect. The Lorentzian splitting surface  $\gamma$ present in the replica wormholes similarly contributes a finite action \cite{Louko:1995jw,Colin-Ellerin:2020mva}.

Since these replica wormhole configurations have finite gravitational action,\footnote{Including quantum corrections, we should interpret the gravitational action as a renormalised effective action, whereby a cutoff-dependent contribution from the path integral over fluctuations is cancelled by appropriate counterterms.} we take the position that it is reasonable to include them in the path integral, even if such spacetimes may have properties that are both novel and unfamiliar. This is particularly justified in light of the classic result that smooth configurations have measure zero in the natural sense associated with quantum field theory path integrals.  This result is deeply related to the familiar strong UV effects (and the associated divergences) in standard quantum field theory.  The same logic then suggests that the gravitational path integral must include singular configurations as well, and finite action provides a reasonable principle for which to include.

The claim now is that spacetimes of replica wormhole topology provide new saddles that contribute to $S^{\text{swap}}_n$ in the context of evaporating black holes. For $n>1$ such saddles necessarily require a complex metric, since the replica wormholes with Lorentzian metric as described cannot solve the semiclassical equations of motion at the singular splitting surface. In Euclidean signature, a saddle would require altering the metric to smooth out the conical defect at $\gamma$, but the same smoothing cannot be achieved in Lorentzian signature without complexifying the metric. The need for complex saddles should not surprise or worry us,  since asymptotic evaulation of oscillatory integrals generically requires deforming the contour of integration to pass through a saddle-point in a complexified domain, and such deformations are required to describe any quantum mechanical tunnelling process as a saddle-point contribution to a real-time path integral.

 While the existence and importance of saddles with such topology may seem surprising, and while a full computation at finite $n$ for Lorentz-signature evaporating black holes has yet to be performed, there is strong evidence for this claim from the following known results.  First, one can indeed find saddles with non-trivial splitting surfaces when one imposes similar boundary conditions in contexts where the effects of bulk quantum fields can be ignored \cite{Colin-Ellerin:2020mva}. Second, at least for $n=2$, analogous Euclidean-signature saddles have been constructed that include back-reaction from quantum fields \cite{Mirbabayi:2020fyk}.  Third, formal calculations at non-integer $n$ allow one to study the limit $n\rightarrow 1$.  For $n-1\ll 1$, one finds a replica wormhole saddle precisely when the $n=1$ Hawking saddle contains a so-called Quantum Extremal Surface (QES). In such a case, the effective action (i.e., minus the logarithm of the path integral) exceeds the action of the $n=1$ Hawking saddle by $(n-1)S_\mathrm{gen}(\gamma)$, where $S_\mathrm{gen}(\gamma) = \frac{A(\gamma)}{4G} + S_\mathrm{matter}(\gamma)$ is the generalized entropy of the QES \cite{Lewkowycz:2013nqa,Faulkner:2013yia,Dong:2016hjy,Dong:2017xht}. Here, $A(\gamma)$ is the area of the splitting surface, and $S_\mathrm{matter}(\gamma)$ is the von Neumann entropy of matter fields on $\Sigma_u\setminus \island$ in the state described by the $n=1$ saddle.  Finally, evaporating black holes do in fact contain a non-trivial QES $\gamma(u)$ located very close to the surface where the past light cone of retarded time $u$ intersects the apparent horizon\footnote{Recall that the apparent horizon is the surface where the outgoing null expansion vanishes.} of the evaporating black hole \cite{Penington:2019npb,Almheiri:2020cfm}.

%The existence of this QES may come as a surprise since such spacetimes contain no classical extremal surfaces.  But since the magnitude of this failure is governed by the back-reaction from the stress-energy of Hawking radiation, the associated quantum contributions to $S_{gen}$ can be large enough to produce a QES.

The swap entropy that results from approximating the path integral by the replica wormhole saddle is easy to calculate using the above results for the limit $n\rightarrow 1$.  The result is the `quantum extremal surface formula',
\begin{equation}
	S^\mathrm{swap} = \min \operatorname*{extr}_\gamma S_\mathrm{gen}(\gamma).
\end{equation}
The $\operatorname*{extr}_\gamma$ denotes extremisation of the generalised entropy with respect to variations of the splitting surface $\gamma$, which is the criterion for finding a saddle point; a resulting stationary surface is a QES. If there is more than one QES, $\min$ denotes choosing the QES with the minimal value of $S_\mathrm{gen}(\gamma)$, which follows from choosing the dominant saddle point.

 In this context, we can think of the Hawking wormhole for the swap entropy as a replica wormhole for which the splitting surface is trivially given by the empty set $\gamma = \emptyset$, with $S_\mathrm{gen}(\emptyset) = S_\mathrm{thermal}(u)$ given by the von Neumann entropy of Hawking radiation as computed above, with no area contribution. In contrast, to good approximation the non-trivial QES gives $S_\mathrm{gen}(\gamma(u)) \sim \frac{A(u)}{4G}$, where $A(u)$ is the area of the evaporating black hole apparent horizon at retarded time $u$,  since the entropy of quantum fields on $\Sigma_u\setminus \island$ is relatively small.

It now remains only to minimise over these two possibilities, choosing the dominant saddle between the Hawking spacetimes above and the new replica wormholes. But the two generalised entropies are precisely the two curves shown in figure \ref{fig:Page} that took part in our discussion of the Page curve.  As a result, by selecting the minimal value we find a phase transition precisely at the Page time, and indeed, as advertised we find that the swap entropy is given precisely by the Page curve of figure \ref{fig:Page}.

\section{Baby universes and superselection sectors}

We argued above that $S^\mathrm{swap}$ for multiple copies of the Hawking radiation receives contributions from replica wormholes that cause it to differ from the von Neumann entropy $S\left(\rho^{(1)} \right).$  Such a discrepancy requires the radiation from our various black hole evaporation experiments to be correlated.  But what is the mechanism that produces such correlations, and what form do they take?

The Hawking and replica wormhole saddles suggest a direct answer to this question. It is clear that they naturally entangle the Hawking radiation with whatever degrees of freedom remain inside the black hole and do not escape to $\scri^+$.  This is of course what leads to the usual statement that the state $\rho^{(1)}$ of Hawking radiation on $\scri^+$ is highly mixed.

However, in the replica wormhole saddles the island $\island$ inside a given black hole (in the bra spacetime) can be glued to the island $\island$ inside another black hole (in the ket spacetime).  The result is a modification of the inner product on these black hole interiors, so that we do not have a separate Hilbert space factor for the interior of each black hole, but only a single overall Hilbert space that describes the interiors of all black holes that happened to form.

In part for historical reasons, this Hilbert space is  called the `baby universe Hilbert space'. One motivation for this name can be seen by considering a black hole that evaporates entirely, to leave behind an essentially flat Minkowski spacetime containing Hawking radiation. After the complete evaporation the black hole interior may be thought of as a closed `baby universe', no longer attached to the `parent' asymptotically flat universe from which it was born. While this description requires some assumptions about the physics once the black hole becomes of Planckian size, we emphasise that it is only a pedagogical crutch, and no such assumptions are required for our conclusions \cite{Marolf:2020rpm}.

 Due to the contribution of replica wormholes, the presence of such baby universes may alter the Hawking radiation produced from any given black hole, and in turn an experiment on a black hole may alter the state of baby universes. This dependence on the state of baby universes highlights an implicit assumption in our discussions above, namely a choice of initial state of closed universes. It merely happens that our discussion was conducted in the absence of initial closed universes (called the Hartle-Hawking baby universe state $|\mathrm{HH}\rangle_\mathrm{BU}$ in \cite{Marolf:2020xie,Marolf:2020rpm}, after noting that this is equivalent to having no boundaries besides those explicitly required for the calculation in question). By modifying the initial state of baby universes, we may obtain different predictions for subsequent experiments. Furthermore, such modifications of the baby universe state occur naturally whenever an observation of the sort discussed above is made.

We thus conclude that entanglement between Hawking radiation and the baby universe Hilbert space induces correlations between the widely-separated evaporation experiments required by our swap entropies, explaining the discrepancy between von Neumann entropy and swap entropy. But such correlations which do not decay with spacetime separation would appear to violate cherished notions of locality, perhaps allowing causality-violating transfer of information between replicas. Fortunately,  this should be forbidden by arguments going back to the 1980's \cite{Coleman:1988cy,Giddings:1988cx,Giddings:1988wv}, which limit the dependence between separate experiments to classical correlations.

The essence of this argument is that any observable on a single black hole, formalised by a set of boundary conditions on a single flat asymptotic region (or an identified bra and ket copy of such to form a single boundary of an in-in contour as above), may be considered as an operator on the baby universe Hilbert space \cite{Marolf:2020xie,Marolf:2020rpm}. This encapsulates the idea above that the an observation of Hawking radiation both depends on and affects the state of baby universes. The set of all such operators generates an algebra $\mathcal{A}_\mathrm{asympt}$ of asymptotic observables. Independent observations on multiple black holes are described by a product of such operators, and correlated observations (such as the measurement of the permutation operators described above) are described by a sum of such products. Furthermore, the path integral with a given set of boundary components does not require us to specify any particular ordering of those components. As such, the algebra $\mathcal{A}_\mathrm{asympt}$ is commutative: the predictions for any two experiments on separate black holes does not depend on the order in which they are performed. Using this mutual commutativity,  all the asymptotic observables may be simultaneously diagonalised. This means that the baby universe Hilbert space is spanned by mutually orthogonal simultaneous eigenstates $|\alpha\rangle_\mathrm{BU}$ called $\alpha$-states, in which all single black hole observables take definite values. Hence, if baby universes are prepared in one of the special states $|\alpha\rangle_\mathrm{BU}$, there can be no correlations of the sort described above. In other words, the baby universe Hilbert space is a direct sum of superselection sectors with respect to the algebra $\mathcal{A}_\mathrm{asympt}$ of asymptotic observables.

Correlations between separate asymptotic regions can arise when the baby universe state is in a superposition of many $\alpha$ states, $|\psi\rangle_\mathrm{BU} = \sum_\alpha \sqrt{p_\alpha}e^{i\theta_\alpha}|\alpha\rangle_\mathrm{BU}$. From this, the expectation value for any asymptotic observable is given by an average over the definite values in each superselection sector, weighted by  $p_\alpha = |\langle\alpha|\psi\rangle_\mathrm{BU}|^2$. But this means that all correlations can be described by a classical probability distribution, with no interference between different superselection sectors, and no dependence on the relative phases $\theta_\alpha$. This is in particular true for the Hartle-Hawking baby universe state $|\psi\rangle_\mathrm{BU}= |\mathrm{HH}\rangle_\mathrm{BU}$, corresponding to an empty set of initial closed universes.

%  And indeed, while we will not go into details here, formal properties of the path integral require that baby universe Hilbert space decompose into so-called superselection sectors (aka $\alpha$-sectors) \cite{Marolf:2020xie,Marolf:2020rpm} with respect to the algebra ${\cal A}_{asympt}$ observables at the asymptotic boundaries.

%This perhaps technical-sounding statement simply means that there is a complete set ${\cal C}_{BU}$ of (diagonalizable) mutually-commuting baby universe observables that also commute with {\it all} elements of ${\cal A}_{asympt}$.  Thus the eigenvalues $\{\alpha \}$ of ${\cal C}_{BU}$ cannot change in any process defined at the asymptotic boundaries.  And in particular there can be no interference terms between eigenstates of ${\cal C}_{BU}$ with distinct eigenvalues $\{\alpha \}$.  As a result, given any initial baby universe state $|\psi\rangle_{BU}$ with amplitudes $\langle \{\alpha \} | \psi \rangle_{BU}$ for each set of eigenvalues $\{\alpha \}$, the results are equivalent to having a mixed state defined by an ensemble of eigenstates $|\{\alpha \}\rangle$ with probabilities $|\langle\{\alpha \} | \psi\rangle_{BU}|^2$. This is in particular true for the Hartle-Hawking baby universe state $|\psi\rangle_{BU}= |HH\rangle_{BU}$, corresponding to an empty set of initial closed universes.

\section{Conclusions}

We may now interpret our calculations of the swap entropies in the light of the general comments regarding baby universes. Both the original Hawking evaporation computation and the calculation of swap entropies above were performed in the baby universe Hartle-Hawking state.  So both results should be interpreted as an average over an ensemble of superselection sectors labelled by orthonormal baby universe states $|\alpha\rangle_\mathrm{BU}$ with weights given by $|\langle\alpha|\mathrm{HH}\rangle_\mathrm{BU}|^2$.

The swap entropy in any given superselection sector is equal to the von Neumann entropy in that sector, since passing to a specific superselection sector eliminates correlations and hence \eqref{eq:uncorrelatedSwap} applies. We interpret the swap entropy obtained from the replica wormholes as an average over the von Neumann entropy of Hawking radiation in each sector, and conclude that in typical sectors, this entropy follows the Page curve.

In contrast, the computation of the entropy from the Hawking wormholes alone is interpreted as the von Neumann entropy of the average state. This may be much larger than the average entropy, since von Neumann entropy is not a linear observable. The key point here is just the familiar statement that an ensemble of pure states is generically highly mixed. But since any particular asymptotic observer will obtain experimental outcomes consistent with a single superselection sector, this large entropy can never have observable consequences.

%Furthermore,
% since ${\cal C}_{BU}$ is a complete set of baby universe observables, there can be only one baby universe state for each $\{\alpha \}$.  So in a given superselection sector there is no room for entanglement with baby universes, and our black hole evaporation experiments must become uncorrelated. Thus for each $\{\alpha \}$  the swap entropies {\it do} in fact describe the von Neuman entropies of the Hawking radiation $\rho^{(1)}_\alpha$ emitted by any given black hole; i.e., Hawking radiation in a given superselection sector satisfies BH unitarity in a naive sense!  But this remains consistent with Hawking's original calculation, which we now understand to compute the ensemble-average of $\rho^{(1)}_\alpha$.  The key point here is just the familiar statement that an ensemble average of pure states is generally highly mixed.

Furthermore, in appropriate contexts additional formal arguments require the density of states in the asymptotically-flat sector of the theory associated with any given superselection sector to be bounded by the Bekenstein-Hawking entropy \cite{Marolf:2020xie}. This guarantees (under the relevant assumptions) that the Page curve for swap entropy bounds the entropy of Hawking radiation in every superselection sector.

In summary, baby universes and superselection sectors are natural consequences of the bulk gravitational path integral whose existence provides sufficient physics to allow both BH unitarity and the original Hawking evaporation calculation to hold without substantial modification.  The calculation of the Page curve from replica wormholes can be interpreted as directly supporting this paradigm. This is achieved without introducing fundamentally new physics, requiring only a conservative interpretation of well-established physical laws. This same physics appears to also resolve at least certain other long-standing issues in quantum gravity \cite{Hsin:2020mfa,Chen:2020ojn}, though there is certainly much more to learn.  Again, we refer the reader to the literature \cite{Penington:2019kki,Almheiri:2019qdq,Marolf:2020xie,Almheiri:2020cfm,Marolf:2020rpm} for more complete discussions of both related physics and open issues. In particular, our brief introduction above should suffice to follow the discussions in sections 7.2 and 7.3 of \cite{Marolf:2020rpm} without studying the rest of that work in detail.

%It is of course no surprise that the radiation from each black hole is correlated with {\it something}.  Indeed, the pure state with which we begin can evolve to a mixed state on $\scri^+$ only through entanglement with some other part of our system.   And we would expect $\scri^+$ to become entangled with the interior of our black hole.  The interesting fact about replica wormholes is that gluing together the black hole interiors as described above leads to correlations among the Hawking radiations on different copies of $\scri^+$.

%\HM{New contribution to $S^\mathrm{swap}$. But no change to Hawking density matrix. How to reconcile?}
%\DM{Keep this short.  Clearly requires correlations.  But path integral must formally give a pure state, so this must involve entanglement with some not-yet-discussed subsystem.  This is the baby universe sector!  Entanglement forms from stuff disappearing into each black hole.  Mention that there is a formal argument that BUs only lead to superselection sectors (but do not describe the argument here).  Also mention that there is an argument for a bound on the density of states in each superselection sector.  }

\acknowledgments

\paragraph{Acknowledgements} We thank Bob Wald for conversations motivating us to write this overview.   We also thank Sean Collin-Ellerin, Xi Dong, Steve Giddings, Mukund Rangamani, Douglas Stanford, Zhencheng Wang and many others for useful conversations about baby universes and superselection sectors.  We are grateful for support from NSF grant PHY1801805 and funds from the University of California. H.M.~was also supported in part by a DeBenedictis Postdoctoral Fellowship.

\bibliographystyle{JHEP}
\bibliography{biblio}

\end{document}